\documentstyle[revtex,eqsecnum]{aps}

\begin{document}
\draft
%
\section   { COVER PAGE}
 
\section {REVTEX compuscript}

  \vfill\eject
 
\begin{title}
  FREQUENCY SEPARATION METHOD  FOR  RELAXATION PROBLEMS \\

\end{title}

\author{Spiros Alexiou} 
\begin{instit}
 Physique Atomique dans les Plasmas Denses, Universite Paris VI\\
 T22-23, Case 128, 4 Place Jussieu, 75252 Paris Cedex 5\\
 and\\
LULI,
 Ecole Polytechnique, 91128 Palaiseau Cedex, France
 \\
\end{instit}
\receipt{ 11 October 1995}
\begin{abstract}
 The  modern era in spectral line broadening began with the understanding that
 the slow(quasistatic) ion and fast(impact) electron perturbers could be 
 treated separately. The problem remained of unifying these two theoretical
 limits. A scheme for this unification is presented here that has at its 
 foundation a fundamental observation that is supported by analytical theory
 and is further demonstrated by computer simulation. The fundamental 
 observation  is that the ions and electrons can be separated most of the
 time, and that a frequency separation within each perturber subsystem can be
 used for unification. That is, the rigorous inclusion of slow, but not 
 necessarily static ions together with the correct impact ion perturbations,
 will produce valid ionic line shapes. We show that a frequency separation
 may be effected to exactly include the fast modulation limit in a variety of
 modern methods that can deal with the intermediate regime between the fast 
 and slow frequency limits of the perturbation.

 Published in Phys.Rev.Lett{\bf 76}, 1836(1996)
 
\end{abstract}
 \narrowtext
 
 
 A large variety of systems with relaxation behavior\cite{gupta,strat} 
  can be described in terms of an active subsystem in interaction
 with an
 external  \lq\lq heat bath"\cite{Fano,Kubo}.
 The interaction with the heat bath is usually taken to be a 
 stochastic, time-dependent potential  V(t) that does $not$ depend on the
 subsystem of interest. This interaction
   causes a
 decoherence, or memory loss, described by a set of  time evolution
  operators that
  have, after a characteristic time, essentially random phase differences
  for different perturber configurations\cite{flor}. In
  the following, spectral line
 broadening will be examined as an example of this behavior. For this case,
 two limits have been much studied. These are usually referred to as the Kubo
 fast(FML) and slow(SML) modulation limits. The technique to be presented here,
 in fact, will be shown to be more widely 
 applicable\cite{Kubo}.  
 
 The  purpose of this letter is to propose a frequency separation for
 the treatment of the intermediate regime between the fast and
 slow modulation limits in the context of spectral line broadening in plasmas.
 Obtaining a theory that encompasses this  intermediate regime  is critically
 important in a number of physical problems, including Stark broadening,
 since the SML is attained extremely slowly as the fluctuation frequency is
 decreased. With the frequency separation technique(FST) proposed here,
  practically all existing methods for the treatment of the intermediate
 regime overcome their problems. Of great practical importance for
 plasma spectroscopy is that the FST, in conjuction with the recently 
 developed 
  Frequency Fluctuation Model(FFM)\cite{FFM,Talin}
  results in an $ultrafast$ method that is
 valid for $all$ $parameter$ $ranges$ and $all$ $kinds$ of emitters.

The FST idea comes from the following observation: The line profile is
 determined by V(t). V(t) is due to $both$ electrons and ions, yet in line
 broadening calculations one almost always calculates their contributions
 separately. It must be emphasized that $there$ $is$ $no$ $mathematical$
 $justification$ $for$ $this$ $step$: The usual $intuitive$ explanation is
 that for the vast majority of practical cases, either electrons or ions
 dominate the broadening, so that this separation should not introduce a 
 large error\cite{Seidel0}. In fact, several
  tests of the electron-ion separation in
 a parameter range where electron or ion broadening was dominant have
 given excellent practical results. In the present letter we further give
 a general formal proof for this common belief by means of a
 $Gedankencalculation$.

There remains the question of what happens when electron and ion broadening
 are comparable. Therefore to illustrate the separation hypothesis for
 comparable electron and ion broadening effects, a test calculation
 of the $H_\alpha$ line emitted from a plasma with electron density of
 $4$x$10^{17}cm^{-3}$ and temperature of 1 eV, has been performed. Fig.1
 presents the
  the dipole autocorrelation function, C(t),
 defined as 
 the Fourier transform of the line profile, resulting from a 
 calculation of the emitting atom perturbed by 
  a)electrons only(dashed-dotted), b)protons
 only(dotted) and c) electrons and protons(solid line).
     All calculations
 involve simulations performed with the non-interacting particle
 collision time-statistics method\cite{Kesting} using 1000 configuration
 averages, at which point convergence was achieved.
 The Gigosos et.al. method\cite{Gig} has been
 used for solving the 
Schr\"odinger equation
  and the code has been benchmarked against standard  
calculations\cite{Kesting,Seiaust}. The electron and ion broadening
 contribute almost equally to the line shape as is evidenced by their
 individual C(t). The product of these individual correlation
 functions is represented by the dashed line, and it can be seen that it is
 in good agreement with the correlation function calculated with both
 electrons and ions. This calculation indicates that the 
 electron-ion separation is  a valid assumption 
$even$   in those rare cases where  ions and electrons have 
 comparable  
 contributions, and where the separation would seem to be least
 justified.

\underline{Fig.1 goes here}

In a unified treatment, that views electrons and ions on an equal footing,
 the main difference between the ionic and electronic fields is their 
 fluctuation frequency. In this regard, the electron-ion separation is, in
 effect, a frequency separation of the total field into fast and slow(not
 necessarily quasistatic) 
 components.

There are two implications in the electron-ion separation: First,
 $to$ $the$ $extent$ $that$
  electrons may be treated in the impact approximation(IA),
 and to the extent that the IA is a solved problem, their part is
  $solved$ and we only need at the end to convolve the ionic and electronic
 profiles. This is the main reason for neglecting electrons in our discussion
  and focusing on the ionic part from now on. 
 In situations of practical interest, the ionic dynamical behavior can
 range from the ion impact to the quasistatic regime and it is
 important to have methods capable of correctly treating ion dynamics.
 A particularly promising modern method is the Frequency Fluctuation Model
 (FFM)\cite{FFM,Talin}. This is an ultrafast stochastic  method capable of
 treating ion dynamical effects, whose  usefulness for the analysis 
 of todays highly complex experiments involving plasma spectroscopy cannot
 be overemphasized. It is in conjunction with the FFM that the frequency
 separation proposed here should have its biggest impact in plasma
 spectroscopy. 

The second implication of the electron-ion separation is that if we were 
 allowed to decompose the total electron-ion field into a fast electron part
 and a slow ionic part, can we do the same with the ion field alone? In
 other words, if $C_{if}(t)$ and  $C_{is}(t)$  denote the 
 autocorrelation functions(AF) due to the interaction of the emitter with
 the fast and slow parts of the ionic field respectively(we specify
 later on what constitutes a fast or a slow component), and $C_i(t)$
 is the ionic AF, can one write
\begin{equation}
C_i(t)
\approx
 C_{if}(t)C_{is}(t)
\ 
\ 
\    ? 
\end{equation}

We proceed to answer this question by means of the following
 $Gedankencalculation$: Consider an arbitrary stochastic interaction V(t),
 divided into fast and slow components and further suppose that for each
 particular realization 
(thereafter called configuration) of
 the interaction, either the fast or slow component will be dominant. In such
 a case, suppose that in a large number N of configurations we have $N_s$
 of them that are dominated by the slow component and $N-N_s$ dominated by the
 fast component. Let $Y_k(t)$ denote the contribution to C(t) of
 the kth configuration, i.e. $C(t)=
\sum^N_{k=1}
Y_k(t)/N$.
 The weak component in configuration k  will have a 
 $Y_k(t)
\approx 1$. Consequently one obtains
\begin{equation}
C(t)\approx {C_1(t)+C_2(t)\over N}  
\end{equation}
 where $C_1(t)=\sum_{k=1}^{N_s} Y_k^s(t)$ and 
 $C_2(t)=\sum_{k=N_s+1}^N Y_k^f(t)$, with the superscipt s or f meaning
  that Y has been computed as a result of the action of the slow or
 fast component alone, respectively. Similarly, writing
\begin{equation}
C_f(t)={N_s+\sum_{k=N_s+1}^N Y_k^f(t)\over N} 
\end{equation}
 and \begin{equation}
C_s(t)={N-N_s+\sum_{k=1}^{N_s} Y_k^s(t)\over N}, 
\end{equation}
 one may show that
 the absolute error, $C(t)-C_s(t)C_f(t)$ 
  is $\lbrack C_s(t)-1\rbrack\lbrack C_f(t)-1\rbrack$. 
 This result justifies the separation approach in two cases:
 the case
 where the fast or slow component is dominant and also the case of short
 times, where C(t) is close to unity  and the relative error is small,
 in agreement with the results of the numerical simulations.
   The assumption in this derivation is that configurations
 where $Y_f$ and $Y_s$ are comparable are rare and that in the vast
 majority of configurations either the fast or slow component
 dominates. This is  confirmed by detailed simulations and is
 also expected physically for any distribution function that is not
 too narrow.  We further note that the argument given by Stamm, Smith
 and Talin\cite{SST}, that the slow component is essentially static during
 a fast collision and that one can neglect the entanglement of the fast
 and slow collisions applies equally to the fast-slow separation
 involving only the ion perturbation, as above. 

This $Gedankencalculation$ is in fact quite general and applies both to the
 electron-ion separation and to the fast-slow separation of the ion
 component.  In a practical test\cite{flor} 
 of the frequency separation on the $ionic$ $component$ $alone$,
 Eq.(1) was very well satisfied; however, this test was done for a parameter
 range where the slow component was dominant. There remains the question of
 what happens when the fast and slow components are comparable, and this is
 a more important question in the context of the fast-slow ion field
 separation, because there is no neutrality constraint relating the fast
 and slow components, as in the electron-ion separation.  
  That is, it is clear that the FST is correct at short times and when
 either the fast or slow component is dominant; however, it is
 less clear when these components are comparable.
 Although smoothness considerations imply that the error introduced by
 the FST should be small, this proposition must be confirmed by a
 calculation. Fig.2 illustrates the accuracy of the FST applied to the
 ion broadening in a parameter range where the fast and slow ion 
 components are comparable, i.e. Lyman $\alpha$ transition from a plasma of
 singly ionized argon perturbers in a 1.38 eV,
 1.2x10$^{16} cm^{-3}$ plasma.  Using the same methods
 as for the calculation in Fig.1, we have calculated in Fig. 2 the
 autocorrelation function C(t) due to   
 a) the fast ionic field component only(dotted line), b) 
  the slow component only(dashed) and c) both fast and
 slow component(i.e. the net ionic) (solid line). Also shown is the
 product of a) and b)(dash-dotted line). All simulation calculations
 involve 1000 configuration averages, at which point convergence was
 achieved.  We note once again that Eq.(1) is very well satisfied.
 As a final remark, for this particular case electrons were
 neglected. Their 
 broadening is almost an order of magnitude less but  can, of course, be
 included by profile convolution.

\underline{Fig.2 goes here}

We have thus far said nothing on what constitutes  a \lq\lq fast" or \lq\lq
 slow" component. The key point is that the fast component must be chosen
 so that it $is$ treatable by the IA. 
  The IA is valid\cite{griem2} for perturbers with 
   $v/\rho\gg HWHM$, where $v$ and $\rho$
 are the particle velocities and impact parameters respectively  
   and  HWHM denotes the half width at half maximum. 
  The way to \lq\lq chop off" the fast component is then 
 to  determine a $v/\rho=\Omega\ge HWHM$; perturbers involving higher
 frequency components(i.e. higher $v/\rho$) belong then to the fast
 component. Technical complications, associated with the precise
 meaning of the $\ge$ in the above 
  equation\cite{Kestaustin}
   are deferred to a longer paper, as are practical ways to
 estimate the HWHM from the width of the impact component.
 
 At this point it should be stressed that we chose to separate out the
 impact fast component, not only because the impact limit is a $known$
 limit, but also because slower, non-impact components are correlated
 and may $not$ be frequency-separated: Only 
 components varying rapidly on the memory
 loss time scale may be separated. An important implication of this fact
 is that at high densities a nonnegligible part of the electrons is
 no longer impact and becomes correlated with the ionic field.

 We now come to the implications of the frequency separation for
 line broadening calculations: With the frequency separation
 the ionic calculation consists of computing the impact
 contribution(fast part) and the slow part contribution. It is clear that
 in this way the ion impact limit is recovered $exactly$ $and$ $in$ $as$
 $sophisticated$ a $form$ $as$ $desired$: For example, one may use the
 perturbative IA\cite{griem2},  the $exact$ semiclassical dipole 
  results\cite{Pfennig} (for hydrogenic emitters) or  fully non-perturbative
  results including higher multipole 
  effects\cite{Alexiou}.
  Further, this
 separation of the ionic field into fast(impact) and slow components
 improves the performance and/or reliability of $every$ method capable
 of dealing with the intermediate(ion dynamical) regime, which will now
 only have to correctly compute the contribution of the $slow$ part:
 For simulations\cite{Stamm}, it is very expensive  to have to integrate
 over sharp peaks. Such peaks are due to fast perturbers, which are
 correctly included in the impact part and do not appear in a simulation
 of the slow part. For the collective coordinate method\cite{flor},
 this separation is very important for the efficiency of the method,
 which would otherwise have to solve large linear systems.
 Finally, stochastic methods have problems in attaining the ion impact
 limit. In the case of  the -now obsolete- 
 Model Microfield
 Method(MMM)\cite{MMM},
  which may only be valid  for neutral hydrogen lines, the (perturbative
 dipole) impact limit is recovered with a 
 delay, even for hydrogen. In the
 case of the  
   FFM, the impact limit is $not$\cite{Talin}
  a limiting case of the formalism, since
 it models the fluctuations of the interaction through a Stark
 component mixing model. 
  Finally, experience with the   
  BID\cite{Dufty} method's approach to the impact limit is limited, but
 again the best that can be attained is the perturbative dipole
 impact limit.
 
   In the FFM (and also BID and  MMM) all the time variation of
  the microfield  is modelled via the covariance of the electric field
$\Gamma_{EE}(t)=
 \langle {\mathbf E}(t){\mathbf\cdot }{\mathbf E}(0)\rangle$.
 With the frequency separation, it is only the slow part of this
 covariance that will appear in the FFM(or BID); the fast part $has$ $been$
 $chosen$ so that it is treatable by the IA.    
  
As an example, the FST will be illustrated within the context of the
 FFM.
  The FFM will be used to describe only the slow component by
 employing a covariance without the rapidly fluctuating component.
 For example, to
  describe the  $\Gamma_{EE}$(t) that represents only the slow component
 of the interaction between neutral emitters and independent perturbers
 moving along straight trajectories\cite{Brissaud2}, the upper limit of the
 velocity integration can be replaced by a finite value, $\Omega\rho$,
 with $\rho$ the impact parameter and $\Omega$ the frequency dividing the
 fast and slow components. This results in the expression for, $\lbrack
 \Gamma_{EE}(t)\rbrack_{slow}$, the slow part of the ion field covariance:
\begin{equation}
\Gamma_{slow}(t)=2\pi n({Ze\over 4\pi\epsilon_0})^2\int dv{ f(v)\over vt}
\lbrack F(Q,b)-F(Q,-b)\rbrack\end{equation}
where $Q={v\over\lambda_D\Omega}, b=vt/\lambda_D$ and
\begin{equation}
 F(Q,b)=\int_Q^\infty dy e^{-y}e^{-\sqrt{y^2+b^2+2b\sqrt{y^2-Q^2}}}
\end{equation}
 with $\lambda_D$ the Debye length, n the density, $f(v)$ the velocity
 distribution and Z the ion charge.

   It must be kept in mind that this result is given merely as an
 illustration, since more sophisticated models for $\Gamma_{EE}(t)$
 may be used, either analytic\cite{Berkovsky}, or numerical.      
 For example, if the covariance is determined from
 simulations, one should exclude the contribution to the total
 field of the particles giving rise to a fast component.

Summing up, the frequency separation proposed in this work has solid
  theoretical as well as practical support and should prove very
 useful in improving existing methods dealing with the intermediate
 regime. The FST has been shown to work when either the fast
 or slow component is dominant and smoothness arguments, as well as
 practical tests, indicate
 that it will indeed work for all cases of interest.
 The most important $practical$ application is that 
  with the frequency  
 separation coupled to the FFM we have an ultrafast robust method, 
  valid for all parameter ranges, a development of great importance
 for plasma spectroscopy.  We note that the incorporation of the FST
 into the, now obsolete, MMM could furnish a unified profile for
 neutral emitters only, while the combination of the FFM and FST
 will provide a unified formulation for all emitters, neutral or
 charged for all plasma conditions. This now moves the focus of the
 outstanding problems to the computation of the strong
 interactions to the impact limit, which is a subject of much current
 interest\cite{Alexiou,Gunter,Oks}.

 Part of this work was done at PIIM, Equipe Diagnostics dans les gaz et des
 Plasmas and I'd like to thank R.Stamm and B.Talin for useful discussions
 and the referees for 
 useful comments on the manuscript.
 This work was partially supported by
  EEC  contract ERBCHRXCT930377.

\section{Figure Captions}
 
FIG.1. Test of the independent
 electron-ion calculation at a parameter range
 where electron and ion broadening
  are about equally important. 
 
FIG.2. Test of the frequency separation on the ionic broadening.
 Particles with $v/
 \rho >
 \Omega=1.44$x   
 $10^{12}$Hz have been included in the fast component and
 the rest in the slow component. The choice of 
 $\Omega$ was made by equating the perturbative expression for the HWHM
 with one tenth of  
 $\Omega$.

 \enddocument